\documentclass[conference,twocolumn,10pt,a4paper]{IEEEtran}

\usepackage[cmex10]{amsmath} 
\usepackage{amssymb,amsthm}
\usepackage{cite}
\usepackage{tikz} 

\usetikzlibrary{decorations.pathreplacing}
\usetikzlibrary{positioning}
\usetikzlibrary{calc}
\usepackage{pgfplots}
\pgfplotsset{compat=1.9}

\newtheorem{theorem}{Theorem}
\newtheorem{lemma}{Lemma}

\newcommand{\ie}{{\it i.e.},\ }
\newcommand{\eg}{{\it e.g.},\ }
\newcommand{\cf}{{\it cf.} }

\newcommand{\EE}{\ensuremath{\mathbb{E}}}

\begin{document}

\title{A Pseudo-Bayesian Approach to Sign-Compute-Resolve Slotted ALOHA}

\author{
\IEEEauthorblockN{
Jasper Goseling\IEEEauthorrefmark{1},
\v Cedomir Stefanovi\' c\IEEEauthorrefmark{2} and
Petar Popovski\IEEEauthorrefmark{2} 
}
\IEEEauthorblockA{
\IEEEauthorrefmark{1}
 Stochastic Operations Research,  University of Twente, The Netherlands\\
 j.goseling@utwente.nl
}
\IEEEauthorblockA{
 \IEEEauthorrefmark{2}
 Department of Electronic Systems,
 Aalborg University, Denmark\\
 \{cs,petarp\}@es.aau.dk
}
}

\maketitle

%
%
%
\begin{abstract}

Access reservation based on slotted ALOHA is commonly used in wireless cellular access.
In this paper we investigate its enhancement based on the use of physical-layer network coding and signature coding, whose main feature is enabling simultaneous resolution of up to $K$ users contending for access, where $K \geq 1$.
We optimise the slot access probability such that the expected throughput is maximised.
In particular, the slot access probability is chosen in line with an estimate of the number of users in the system that is obtained relying on the pseudo-Bayesian approach by Rivest, which we generalise for the case that $K>1$.
Under the assumption that this estimate reflects the actual number of users, we show that our approach achieves throughput $1$ in the limit of large $K$.
\end{abstract}

%
%
%
\section{Introduction}

Enhancements and redesign of traditional random access algorithms have been in research focus in recent years, instigated by the need for massive and uncoordinated access pertaining to applications and services of Internet-of-Things and, specifically, Machine-to-Machine (M2M) communications.
In particular, it was shown that the performance of random access algorithms can be significantly improved using novel concepts, such as coded random access \cite{PSLP2014}, physical-layer network coding (PLNC) \cite{cocco2011vtc,censor2012bounded,goseling13ita,goseling14massap,GSP2014}, and compressive sensing \cite{CS-MUD,WJW2014,JSBPD2014}.

In this paper we address a typical M2M scenario with a large number of users that sporadically have data to transmit to a single receiver (\eg base station).
The users with data packets achieve communication in two phases, the first consisting of the access reservation and the second of the data transmission.
Our interest lies only in the access reservation phase, in which the users contend to obtain scheduled resources (\ie link time and frequency) for the actual data transmission.
Particularly, we investigate access reservation based on slotted ALOHA, which is a common feature of cellular standards \cite{TS44_060,3GPPTS36.321}.

Our contribution consists of an access mechanism that exploits PLNC and signature codes, as introduced in \cite{goseling14massap,GSP2014}.
The key characteristic of our scheme is that simultaneous transmissions by more than one user does not necessarily lead to a wasteful collision.
Instead, the concept of collision is generalised such that up to $K$ simultaneous transmissions are instantaneously resolved, where $K$ is a design parameter.
If more than $K$ users are simultaneously transmitting, the receiver declares an unresolvable collision and all involved users attempt retransmissions in later slots.

The ability to simultaneously resolve up to $K$ contending users impacts the design of the access protocol parameters.
In this paper we investigate the optimisation of the slot-access probability (\ie the probability with which the contending users attempt access reservation), such that the expected throughput (\ie the expected number of users resolved per slot of the access reservation phase) is maximised. 
We base the investigation on the pseudo-Bayesian approach, as introduced by Rivest~\cite{Rivest1987,bertsekasgallager}.

While in the standard ALOHA model the packets transmitted by users represent atomic units of communication for which the internal structure is hidden, in our case we use a more elaborate model in order to assess the cost introduced by the use of signatures.
Specifically, the ability to simultaneously resolve up to $K$ users impacts the length of the signatures (\ie the user identifiers/addresses), where larger $K$ implies longer signatures and, thereby, longer slots in the access-reservation phase.
In turn, this implies larger number of new arrivals during a slot and a potential rise in the number of the contending users in the subsequent slots.
Nonetheless, the main technical contribution of this paper is that we demonstrate that the throughput of the proposed scheme is increasing in $K$, approaching one for $K\to\infty.$

The combined use of PLNC and the way that signatures are constructed enables the receiver to detect the number of transmitting users in a slot.
This information is used to update the access probability, which is fed back to the users and employed for the contention in the next slot.
Given that this knowledge of the `collision multiplicity' is available, it is well known that throughput one can be achieved~\cite{P1981}, but operating the corresponding protocol requires exponential computational complexity \cite{Ruszinko1997}.
The contributions of this work consist of proposing a simple scheme that achieves such throughput.

Further, the combined use of PLNC and signature coding for random access in a tree splitting framework was the topic of our previous work~\cite{GSP2014}.
In contrast, the scheme proposed in this paper, as any slotted ALOHA-based scheme \cite{R1975}, attempts to resolve collisions simply by retransmissions that are performed in later slots independently by all backlogged users. 
This significantly simplifies the protocol operation, as there is no need for the implementation of state variables that govern the access mechanism at the user side \cite{Massey}.



Finally, we note that suggested approach bears similarities to the multipacket reception-based random access protocols, \cf \cite{ghez88mpr,TZM2001}. The typical assumption in these works is that the signal-separation capability at the receiving end is probabilistic in nature and depends on the number of the simultaneously attempted transmissions.
In contrast, the scenario analysed in this paper could be categorised as random access with perfect multiple reception of up to $K$ packets.
A related comparison between multipacket reception-based protocols and physical-layer network coding based schemes is provided in~\cite{goseling13ita}.

The rest of the text is organised as follows.
Section~\ref{sec:model} introduces the system model.
Section~\ref{sec:strategy} deals with the proposed access strategy.
In Section~\ref{sec:results} we state the main results, whose proofs and the supporting analysis are performed in Section~\ref{sec:analysis}.
Finally, Section~\ref{sec:discussion} concludes the paper.

%
%
%
\section{Model} \label{sec:model}

We start with a general overview.
Communication takes place over a time-slotted channel.
Each user (terminal) sends a \emph{reservation packet} (referred to as a \emph{packet} for brevity), that does not contain any data, such that the objective of the common receiver in this random access process is to learn the identities of the users that are trying to send reservation packets.
Each packet is unique, designed such that the receiver is capable to learn the identities of up to $K$ simultaneously transmitting users, where $K \geq 1$.
The receiver is also able to learn the number of simultaneous transmissions in a slot, denoted by $C$, also when $C > K$.
The above functionalities are achieved through combined use of PLNC and signature codes, as detailed in the sequel.
If $C \leq K$ users transmitted in a slot, their identities become resolved and they depart from the system.
If $C > K$, the users that transmitted simultaneously are not resolved and become backlogged, retransmitting in some later slot.
Further, there is an immediate and perfect feedback after every slot $t$, through which the receiver informs the users about the observed multiplicity $C_t$, and through which the users also learn the access probability valid for the next slot $q_{t+1}$.
Next, we describe each of the elements of our model in more detail. 

\subsubsection*{Slotted time}
Time in the system is slotted and all users are slot synchronised, such that at the point of reception all transmitted packets are received aligned with the slot duration.

\subsubsection*{Arrival model}
There are $M$ users in the system.
Packets, \ie reservation requests, arrive at the users according to a (continuous-time) Poisson process with a sum arrival rate $\lambda$.
We assume that all arrivals occur at users that are not currently contending; under this assumption our model is equivalent to an infinite population model that is often used in studies on ALOHA \cite{bertsekasgallager}.



\subsubsection*{Access mechanism}
All newly arrived packets are immediately backlogged, \ie their transmissions are attempted in the same way as for the packets that experienced collisions previously.{\footnote{The model of the backlog is elaborated in details in Section~\ref{sec:strategy}.}
The transmission of backlogged packets in slot $t$ is attempted with probability $q_t$, which is equal for all users/packets.
The number of simultaneously transmitted packets in slot $t$ is denoted by $C_t$, $t \geq 0$.


\subsubsection*{Transmitted and received signals}
We consider a Gaussian multi-access channel with unit channel gains.
In particular, in slot $t$ the receiver observes
\begin{align}
\label{eq:Y}
Y_t = \sum_{m \in \mathcal{C}_t} X_m + Z_t,
\end{align}
where $Y_t$ and $X_m$ are channel output and inputs, respectively, $\mathcal{C}_t$ is the set of users transmitting in slot $t$, $ | \mathcal{C}_t | = C_t$, and $Z_t$ is is the Gaussian noise with unit variance.

\subsubsection*{Physical-layer network-coding}
We base our approach to PLNC on the compute-and-forward framework~\cite{nazer11compforw}.
The best known achievable PLNC rate for channel~\eqref{eq:Y} is $\frac{1}{2} \log_2 ( 1/ C_t + P )$.
Since $C_t$ itself is unknown at the user side at the time of transmission, we will use rate $\frac{1}{2} \log_2 ( 1/ M + P )$.
The capacity of~\eqref{eq:Y} is  $\frac{1}{2} \log_2 (1+ P )$  and, therefore, a small performance penalty is incurred from the use of PLNC. 
However, as our interest in this paper is in the rate at which users are departing the system, we will not model the impact of the PLNC rate explicitly.
Recall from~\cite{nazer11compforw} that the use of PLNC transforms the considered multiple-access Gaussian channel into noiseless $\mathbb{F}_q$ adder channel~\cite{GSP2014}
\begin{align}
\mathbf{Y}_t = \bigoplus_{m \in \mathcal{C}_t} \mathbf{X}_m,
\end{align}
where $\mathbf{Y}_t$ and $\mathbf{X}_m$, $m \in \mathcal{C}_t$, are $q$-ary sequences representing channel output and inputs in slot $t$, respectively, and $\bigoplus$ is symbol-wise addition in $\mathbb{F}_q$. We assume that field size $q$ is sufficiently large that the result on the computation rate holds.

\subsubsection*{Signature coding}
Each user $m$ is provided with a unique signature codeword $\mathbf{X}_m$.
This implies that the signature length has to be at least $\log_2 M$ bits, but the gist of our signature code approach is that we extend the length of the signatures to enable user identification in the case of up to $K$ simultaneous transmissions.
More precisely, we require that the sums (over $\mathbb{F}_q$) of at most $K$ different signatures are uniquely decodable.
The construction of signatures such that their random sums on adder channel are uniquely decodable has received a lot of attention, \cf \cite{MACbook} for the survey of the known results.
In this paper we use the result from \cite{lindstrom1975,GSP2014}, stating that, given that a sum of up to $K$-out-of-$M$ signatures is uniquely decodable, we can create signatures of length $L$ where
\begin{align}
\label{eq:signature_bound}
L =  \log_2 M \left (K \left(1 - \frac{\log_2 K }{ \log_2 M } \right)^{-1} + 1 \right ) \approx K \log_2 M.
\end{align}
An additional feature of the signature code construction provided in~\cite{GSP2014} is that it provides the receiver with the exact value of $C_t$, also when $C_t > K$.

Obviously, the length is the signatures is proportional to $K$, and, therefore, the length of the slots has to be proportional to $K$.
In the remainder we assume that the length of a time slot is $K$ [s].
Note that this implies that during a time slot the expected number of arrivals is $K \lambda$.
We omit the normalisation with $\log_2 M$, since this is a lower bound on the packet length in any scheme where $M$ users need to be given a unique identifier.

\subsubsection*{Feedback}
After every slot $t$, the receiver feeds back the contending users with the values of $C_t$ and the access probability $q_{t+1}$ that is valid for the next slot. 
If $C_t \leq K$, the users that transmitted in slot $t$ learn that their packets have been successfully decoded and leave the system.
Conversely, if $C_t > K$, the involved users learn that an unresolvable collision occurred and they remain backlogged.
The determination of $q_{t+1}$, which is done by the receiver such that the performance of the access mechanism is optimised, based on the backlog model and the observed $C_t$, is the pivotal contribution of the paper and elaborated in the following sections.

\subsubsection*{Notation}
Let $\Gamma(k,x)$ denote the upper incomplete Gamma function, \ie
\begin{equation}
 \Gamma(k,x) = \int_x^\infty z^{k-1}e^{-z}dz. 
\end{equation}
We will make use of the well known relations:
\begin{align}
\Gamma(k) = \Gamma(k,0)  =  ( k-1)!, \\ 
\sum_{k=0}^{K-1} \frac{x^k}{k!}e^{-x}  = \frac{\Gamma(K,x)}{\Gamma(K)}.
\end{align}

%
%
%
\section{Proposed Strategy} \label{sec:strategy}

Recall from the model description that we ensure that the receiver reliably obtains $C_t$, the number of users that are transmitting.
Therefore, the number of backlogged users in slot $t$, $N_t$, is a random variable whose evolution is
\begin{equation}\label{eq:SecIII_EQN1}
N_{t+1} =
\begin{cases}
A_t + \max\{0, N_t - C_t\}, &\text{if } 0 \leq C_t \leq K,\\
A_t + N_t&\text{if } C_t > K.   
\end{cases}
\end{equation}
where $A_t$ is a Poisson random variable with mean $\lambda K$, referring to to the number of new arrivals (the expected number of arrivals is proportional to $K$, as the length of the packets/slots is proportional to $K$).
The reasoning behind relation (\ref{eq:SecIII_EQN1}) for $N_{t+1}$ is that, in case $0 \leq C_t \leq K$, exactly $C_t$ users are resolved.
If $C_t>K$, all users remain in the system.

In slot $t$ all users independently transmit a packet with probability $q_t$.
The value of $q_t$ is indicated by the receiver at the end of the previous slot, slot $t -1$, and is based on an estimate of the number of backlogged users $N_t$ in slot $t$.
Let $\hat{n}_t$ denote the estimate of the \emph{mean} of $N_t$.
The access probability $q_t$ is chosen as
\begin{equation}
q_t=\min\{\alpha/\hat n_t,1\},
\end{equation}
where $\alpha$ is a constant that depends on $K$ and that will be specified more precisely below.

Next, we turn our attention to the updating rule for $\hat n_t$.
The strategy that we propose for setting $\hat n_{t+1}$ is based on the psuedo-Bayesian algorithm of Rivest~\cite{Rivest1987}, a good treatment of which appears in~\cite{bertsekasgallager}. 
The pseudo-Bayesian approach revolves around the modelling approximation that the number of backlogged users $N_t$ is Poisson distributed.
Based on this assumption, optimal updating rules are devised. 

Rivest~\cite{Rivest1987} considered $K=1$ and $\alpha=1$ and showed that, if $C_t=1$, the a posteriori estimate of the remaining number of users is Poisson distributed with mean $\hat n_t - 1$.
If $C_t > 1$ the a posteriori distribution is no longer Poisson, but it's mean can be expressed analytically and used to update $\hat n_{t+1}$.
It was shown by Rivest through numerical experiments that using this modelling approximation and updating rule leads to a good estimate of $\EE [ N_t ]$ through $\hat n_t$. 
A rigorous proof that the overall strategy is stable is given in~\cite{tsybakov1979ergodicity, tsitsiklis86report}. 

We generalize the psuedo-Bayesian algorithm to $K>1$.
In Section~\ref{sec:analysis} we show that if, $0 \leq C_t \leq K$, $C_t$ users leave the system and the a posteriori estimate of the remaining number of users is Poisson distributed with mean $\hat n_t - \alpha$.
We also show that, if $C_ t >K$, the expected a posteriori number of users is $\hat n_t + C_t - \alpha$.
This provides the following updating rule
\begin{equation}
\label{eq:updatehatn}
\hat n_{t+1} =
\begin{cases}
\lambda K + \max\{0, \hat n_t - \alpha\}, &\text{if } 0\leq C_t \leq K,\\
\lambda K + \hat n_t + C_t - \alpha, \quad &\text{if } C_t > K.   
\end{cases}
\end{equation}
for the estimate of $\EE [ N_{t+1} ] $, see \eqref{eq:SecIII_EQN1}.

%
%
%
\section{Results} \label{sec:results}

The proofs of the results presented in this section are given in Section~\ref{sec:analysis}.

The first result deals with the value of $\alpha$ and the expected number of users that get resolved in a slot, denoted by $\EE[S]$.
We give an exact expression for $\EE[S]$ under the assumption that $N$ is Poisson distributed and that $\hat n = \EE[N]$.
In this section for notational convenience we drop all subscripts $t$.
\begin{theorem} \label{th:SPoisson}
Let $\alpha$ satisfy $\Gamma(K,\alpha) = \alpha^K e^{-\alpha}.$
If $N$ is Poisson distributed and $\hat n = \EE[N]>\alpha$ then $q=\min\{\alpha/\hat n,1\}$ is the access probability that maximises $\EE[S]$. In this case we have $\EE[S]=\alpha\Gamma(K,\alpha)/\Gamma(K)$.
\end{theorem}
Solving the relation $\Gamma(K,a) = a^K e^{-a}$ can be reduced to finding a real root of a $K$-th order polynomial in $a$, which can in general only be done numerically.
In Table~\ref{table:astar} we list the values of $\alpha$ for various values of $K$. 

\begin{table}
\begin{equation*}
\begin{array}{rl}
K & \alpha \\
\hline
1 & 1 \\
2 & 1.61803  \\
3 &  2.26953 \\
4 &  2.94519  \\
5 &  3.63955 \\
6 & 4.34905
\end{array}
\end{equation*}
\caption{Evaluation of $\alpha$ as a function of $K$.}
\label{table:astar}
\end{table}

Next, we turn our attention to the expected throughput  $\EE[T]$, which we define as
\begin{equation}
\label{eq:T}
\EE[T] = \frac{\EE[S]}{K},
\end{equation}
\ie we normalize $\EE [ S ]$ by the slot length, as induced by the length of the signatures.
Computing $\EE [ T ]$ for the values from Table~\ref{table:astar}, which is the maximal expected throughput given $K$, shows that $\EE[T]$ is increasing in $K$, which is also depicted in Fig.~\ref{fig:T}. 
For instance, at $K=2$ is $\EE [ T ] = 0.38$ and at $K=6$ is $\EE [ T ] = 0.53$. 
Note that the maximal expected throughput that can be achieved by the standard slotted ALOHA, \ie when $K=1$, is $1/e\approx 0.37$.
At $K=6$, for instance, we already significantly outperform the standard slotted ALOHA.

\begin{figure}
\centering
\begin{tikzpicture}
\begin{axis}[
height=6cm,
  xlabel=$K$,ylabel={$\EE[T]$}, 
  font=\scriptsize,
]

\addplot[
  line width=.3mm,color=blue, solid,
  ]
coordinates {
(1,0.367879) (2,0.419981) (3,0.457034) (4,0.485595) (5,0.508707) (6,0.528031) (7,0.544574) (8,0.558994) (9,0.571741) (10,0.583139) (11,0.593426) (12,0.602784) (13,0.611356) (14,0.619252) (15,0.626563) (16,0.633363) (17,0.639712) (18,0.645662) (19,0.651255) (20,0.656527) (21,0.661511) (22,0.666234) (23,0.670718) (24,0.674985)
};
\end{axis}
\end{tikzpicture}
\caption{$\EE[T]$ as a function of $K$.}
\label{fig:T}
\end{figure}

Table~\ref{table:ic} shows that the collision multiplicity distribution induced by the optimal access probability is such that the slots with unresolvable collisions are more likely to occur than idle slots for $K > 1$, and this becomes more pronounced as $K$ increases.
Obviously, this shift towards unresolvable collisions is overcompensated by the beneficial effect of an increased throughput.

Our next result deals with the value of $\EE[T]$ for large values of $K$.
In particular, in Section~\ref{sec:analysis} we prove the following:
\begin{theorem} \label{th:Tlim}
\begin{equation}
\lim_{K\to\infty} \EE[T] = 1.
\end{equation}
\end{theorem}
Theorem~\ref{th:Tlim} states that it is possible to achieve a throughput arbitrarily close to $1$ by allowing large $K$, despite the fact that the slot length also increases with $K$, see \eqref{eq:T}.

The above results demonstrate that the throughput performance of the proposed scheme is promising.
However, we have not shown that the approach is stable.
Stability will rely, as in the original pseudo-Bayesian approach, on the fact that the estimate $\hat n$ is following $ \EE [ N ] $ closely enough.
We do not provide a complete proof of stability in this paper.
On the other hand, we do provide in the next section an analysis and discussion of the updating rules for $\hat n$.

%
%
%

\section{Proofs} \label{sec:analysis}

As already outlined, the key assumption of the pseudo-Bayesian approach is that the number of backlogged users is Poisson distributed.
In this section we first establish a number of auxiliary results under this assumption.
For notational convenience, we drop the subscripts $t$ and $t+1$ when no confusion can arise. 
 
 \begin{table}
\begin{equation*}
\begin{array}{ccc}
K & P [ C = 0 ] & P [ C >  K ] \\
\hline
1 & 0.3679 & 0.2642 \\
2 &  0.1983 &  0.2213  \\
3 &  0.1034 &  0.1945 \\
4 &  0.0526 &  0.1756 \\
5 &  0.0263 &  0.1614 \\
6 &  0.0129 & 0.1501
\end{array}
\end{equation*}
\caption{Probabilities of an idle slot $P [ C = 0 ]$ and a slot with unresolvable collision $ P [ C > K ] $ for the optimal value of $\alpha$ as a function of $K$.}
\label{table:ic}
\end{table}

We first provide a result on the a posteriori distribution of the number of users in the system, once it is known how many users have transmitted in a slot.
\begin{lemma} \label{lem:probk}
Let the number of backlogged users $N$ be Poisson distributed with mean $\hat n$.
Suppose that each user is transmitting with probability $q$. Then, the a posteriori probability that there were $n + c$ backlogged users given that $c$ have transmitted is
\begin{equation}
\label{eq:post}
P [ N = n + c | c\ \textnormal{tx} ] = \frac{((1-q)\hat n)^n}{n!}e^{-(1 - q)\hat n}.
\end{equation}   
\end{lemma}
\begin{IEEEproof}
It follows from a thinning argument on a Poisson process $N$ that the probability that $c$ out of $N$ users are transmitting, $c\geq 0$, is
\begin{equation}
P [ c\ \text{tx} ] = \frac{(\hat n q)^c}{c!}e^{-\hat n q}.
\end{equation}
Now, the result follows through an application of Bayes rule:
\begin{align}
P [ N = n + c | & c \text{ tx} ]  = \\
 = & \frac{P [ c \ \text{tx} | N = n + c ] P [ N = n + c ] }{P [ c\ \text{tx} ] } \\
=  & \frac{\binom{n + c}{c}q^c(1-q)^n \frac{\hat n^{n + c}}{(n + c)!}e^{-\hat n}}{\frac{(\hat n q)^c}{c!}e^{-\hat n q}},
\end{align}
which evaluates to \eqref{eq:post}.
\end{IEEEproof}

If $0\leq c\leq K$, all $c$ users that were transmitting become resolved and leave the system.
Therefore, the above result states that in this case the remaining number of users in the system is still Poisson distributed, with the new mean
\begin{align}
\EE[N | c\ \text{tx}] = (1-q)\hat n = \hat n - \alpha,
\end{align}
as reflected in the update rule \eqref{eq:updatehatn} for the case $ 0 \leq c \leq K$.

If there are more than $K$ active users the posterior distribution of the remaining number of users is not Poisson (it is shifted by $c$).
The key idea in the pseudo-Bayesian approach is to approximate and assume that it is still Poisson.
We have
\begin{equation}
\EE[N | c\ \text{tx}] = c + (1-q)\hat n. 
\end{equation}
This mean value is to update the value of $\hat n$ when $c > K$, as  indicated in \eqref{eq:updatehatn}.

We turn our attention to finding the optimal value for $q$.
Let $S$ denote the number of users that gets resolved in a slot.
We have
\begin{equation}
S =
\begin{cases}
c,\quad &\text{if } 0 \leq c \leq K,\\
0,\quad &\text{if } c > K.
\end{cases}
\end{equation}
First we give a result on $\EE[S]$.
\begin{lemma} \label{lem:S}
If the number of backlogged users is Poisson distributed and $\hat n = \EE[N]$,  then the expected number of users that gets resolved is $\EE[S]=q\hat n\Gamma(K,q\hat n)/\Gamma(K).$
\end{lemma}
\begin{IEEEproof}
It follows immediately from Lemma~\ref{lem:probk} that
\begin{align}
\EE[S]
= \sum_{ c = 0 }^K c P [ c \ \text{tx} ] = \sum_{ c = 0 }^K c \frac{(q \hat n)^c }{c!}e^{-q\hat n} =  q\hat n \frac{\Gamma(K,q\hat n)}{\Gamma(K)}.
\end{align}
\end{IEEEproof}

Next, we optimise the access probability $q$.
\begin{lemma} \label{lem:optq}
If the number of backlogged users is Poisson distributed and $\hat n = \EE[N]$,  then $\EE[S]$ is optimized for $q=\max\{\alpha/\hat n,1\}$ where $\alpha$ satisfies
\begin{equation}
\Gamma(K, \alpha) = \alpha ^K e^{- \alpha}.
\end{equation}
\end{lemma}
\begin{IEEEproof}
Immediate from Lemma~\ref{lem:S} by considering
\begin{equation}
\frac{\partial}{\partial q}\EE[S] = \frac{\hat n\Gamma(K,q\hat n)}{\Gamma(K)} - \frac{(q\hat n)^Ke^{-q\hat n}\hat n}{\Gamma(K)}.
\end{equation}
\end{IEEEproof}

We conclude by presenting the proofs of theorems from Section~\ref{sec:results}.
\begin{IEEEproof}[Proof of Theorem~\ref{th:SPoisson}]
From Lemma~\ref{lem:optq}.
\end{IEEEproof}
\begin{IEEEproof}[Proof of Theorem~\ref{th:Tlim}]
Instead of taking $q=\max\{\alpha/\hat n,1\}$ with the optimal value of $\alpha$, we use $q=\max\{\delta K / \hat n,1\}$, where $\delta K$ is an arbitrary fraction of $K$, \ie  $0< \delta <1$.
Then we consider $K \to \infty$ and obtain
\begin{equation}
\lim_{K\to\infty} \EE[T] \geq \lim_{K\to\infty} \frac{\delta K \ \Gamma(K, \delta K)}{K \ \Gamma(K)} = \delta.
\end{equation}
Now, since $\delta$ was an arbitrary value from $( 0,1 )$, it follows that any throughput arbitrarily close to $1$ can be achieved.
Note that in this case the use of a non-optimal $q$ is sufficient to demonstrate optimal throughput for $K\to\infty$.
\end{IEEEproof}


\section{Discussion}
\label{sec:discussion}

We have proposed an access reservation scheme that is based on physical-layer network coding, signature coding and slotted ALOHA type of contention.
Its main features are (i) ability to detection of the number of transmitting users in each time slot, (ii) instantaneous resolution of all transmitting users as long there are not more than $K$, and (iii) optimisation of the slot-access probability, under the assumption that the number backlogged users is Poisson distributed.
As the length of the codewords in the signature codes is proportional to $K$, the length of the slots also scales with $K$, implying increased number of new arrivals during a slot.
Thus, a priori it is not clear how $K$ should be chosen optimally.
Our results demonstrate that throughput is strictly increasing in $K$, and we have shown that throughput one can be achieved.
The most important aspect of our future work will be to rigorously prove stability of the proposed scheme.
Since our scheme closely resembles the pseudo-Bayesian approach of Rivest, we conjecture that stability holds and that techniques similar to those used by Tsitsiklis~\cite{tsitsiklis86report} can be used for the analysis.  

\section*{Acknowledgement}
This work was supported in part by the Danish Council for Independent Research (DFF), grants 4005-00281 and 11-105159.

%
%
%

%
%
%
\bibliographystyle{IEEEtran}

\begin{thebibliography}{10}
\providecommand{\url}[1]{#1}
\csname url@samestyle\endcsname
\providecommand{\newblock}{\relax}
\providecommand{\bibinfo}[2]{#2}
\providecommand{\BIBentrySTDinterwordspacing}{\spaceskip=0pt\relax}
\providecommand{\BIBentryALTinterwordstretchfactor}{4}
\providecommand{\BIBentryALTinterwordspacing}{\spaceskip=\fontdimen2\font plus
\BIBentryALTinterwordstretchfactor\fontdimen3\font minus
  \fontdimen4\font\relax}
\providecommand{\BIBforeignlanguage}[2]{{%
\expandafter\ifx\csname l@#1\endcsname\relax
\typeout{** WARNING: IEEEtran.bst: No hyphenation pattern has been}%
\typeout{** loaded for the language `#1'. Using the pattern for}%
\typeout{** the default language instead.}%
\else
\language=\csname l@#1\endcsname
\fi
#2}}
\providecommand{\BIBdecl}{\relax}
\BIBdecl

\bibitem{PSLP2014}
E.~Paolini, C.~Stefanovic, G.~Liva, and P.~Popovski, ``{C}oded {R}andom
  {A}ccess: {A}pplying {C}odes on {G}raphs to {D}esign {R}andom {A}ccess
  {P}rotocols,'' \emph{IEEE Commun. Mag.}, to appear.

\bibitem{cocco2011vtc}
G.~Cocco, C.~Ibars, D.~Gunduz, and O.~del Rio~Herrero, ``{C}ollision
  {R}esolution in {S}lotted {ALOHA} with {M}ulti-{U}ser {P}hysical-{L}ayer
  {N}etwork {C}oding,'' in \emph{Proc. IEEE VTC 2011 (Spring)}, Yokohama,
  Japan, May 2011.

\bibitem{censor2012bounded}
K.~Censor-Hillel, B.~Haeupler, N.~Lynch, and M.~M{\'e}dard,
  ``{B}ounded-{C}ontention {C}oding for {W}ireless {N}etworks in the {H}igh
  {SNR} {R}egime,'' in \emph{Distributed Computing}.\hskip 1em plus 0.5em minus
  0.4em\relax Springer, 2012, pp. 91--105.

\bibitem{goseling13ita}
J.~Goseling, M.~Gastpar, and J.~H. Weber, ``{R}andom {A}ccess with
  {P}hysical-layer {N}etwork {C}oding,'' in \emph{Proc. Information Theory and
  Applications Workshop, 2013}, San Diego, CA, USA, Feb. 2013.

\bibitem{goseling14massap}
J.~Goseling, ``{A} {R}andom {A}ccess {S}cheme with {P}hysical-layer {N}etwork
  {C}oding and {U}ser {I}dentification,'' in \emph{Proc. IEEE ICC 2014,
  Workshop on Massive Uncoordinated Access Protocols}, Sydney, Australia, Jun.
  2014.

\bibitem{GSP2014}
J.~Goseling, C.~Stefanovic, and P.~Popovski, ``{S}ign-{C}ompute-{R}esolve for
  {R}andom {A}ccess,'' in \emph{Proc. 52nd Annual Allerton Conference},
  Monticello, IL, USA, Sep. 2014.

\bibitem{CS-MUD}
C.~Bockelmann, H.~Schepker, and A.~Dekorsy, ``Compressive sensing based
  multi-user detection for machine-to-machine communication,'' \emph{Trans. on
  ETT}, vol.~24, no.~4, pp. 389--400, Jun 2013.

\bibitem{WJW2014}
G.~Wunder, P.~Jung, and C.~Wang, ``{C}ompressive {R}andom {A}ccess for
  {P}ost-{LTE} {S}ystems,'' in \emph{Proc. IEEE ICC 2014, Workshop on Massive
  Uncoordinated Access Protocols}, Sydney, Australia, June 2014.

\bibitem{JSBPD2014}
Y.~Ji, C.~Stefanovic, C.~Bockelmann, A.~Dekorsy, and P.~Popovski,
  ``{C}haracterization of {C}oded {R}andom {A}ccess with {C}ompressive
  {S}ensing based {M}ulti-{U}ser {D}etection,'' in \emph{Proc. IEEE Globecom},
  Austin, TX, USA, Dec. 2014.

\bibitem{TS44_060}
3GPP, ``{{G}eneral {P}acket {R}adio {S}ervice {(GPRS)}; {M}obile {S}tation
  {(MS)} - {B}ase {S}tation {S}ystem {(BSS)} interface; {R}adio {L}ink
  {C}ontrol / {M}edium {A}ccess {C}ontrol {(RLC/MAC)} protocol},'' {3rd
  Generation Partnership Project (3GPP)}, TS {44.060}, Sep. 2008.

\bibitem{3GPPTS36.321}
{3GPP}, ``{{M}edium {A}ccess {C}ontrol (MAC) protocol specification},'' {3rd
  Generation Partnership Project (3GPP)}, TR {36.321}.

\bibitem{Rivest1987}
R.~L. Rivest, ``{N}etwork {C}ontrol by {B}ayesian {B}roadcast,'' \emph{IEEE
  Trans. Info. Theory}, vol.~33, no.~3, pp. 323?--328, May 1987.

\bibitem{bertsekasgallager}
D.~Bertsekas and R.~Gallager, \emph{Data Networks, 2nd ed.}\hskip 1em plus
  0.5em minus 0.4em\relax Prentice-Hall, Inc, 1992.

\bibitem{P1981}
N.~Pippenger, ``{B}ounds on the {P}erformance of {P}rotocols for a
  {M}ultiple-{A}ccess {B}roadcast {C}hannel,'' \emph{IEEE Trans. Info. Theory},
  vol.~27, no.~2, pp. 145--?151, Mar. 1981.

\bibitem{Ruszinko1997}
M.~Ruszinko and P.~Vanroose, ``{H}ow an {E}rdos-{R}enyi-{T}ype {S}earch
  {A}pproach {G}ives an {E}xplicit {C}ode {C}onstruction of {R}ate 1 for
  {R}andom {A}ccess with {M}ultiplicity {F}eedback,'' \emph{IEEE Trans. Info.
  Theory}, vol.~43, no.~1, pp. 368?--373, Jan. 1997.

\bibitem{R1975}
L.~G. Roberts, ``Aloha packet system with and without slots and capture,''
  \emph{SIGCOMM Comput. Commun. Rev.}, vol.~5, no.~2, pp. 28--42, Apr. 1975.

\bibitem{Massey}
J.~L. Massey, ``\BIBforeignlanguage{English}{{C}ollision-{R}esolution
  {A}lgorithms and {R}andom-{A}ccess {C}ommunications},'' in
  \emph{\BIBforeignlanguage{English}{Multi-User Communication Systems}}, ser.
  International Centre for Mechanical Sciences, G.~Longo, Ed.\hskip 1em plus
  0.5em minus 0.4em\relax Springer Vienna, 1981, vol. 265, pp. 73--137.

\bibitem{ghez88mpr}
S.~Ghez, S.~Verdu, and S.~Schwartz, ``{S}tability {P}roperties of {S}lotted
  {ALOHA} with {M}ultipacket {R}eception {C}apability,'' \emph{IEEE Trans.
  Automat. Contr.}, vol.~33, no.~7, pp. 640--649, Jul. 1988.

\bibitem{TZM2001}
L.~Tong, Q.~Zhao, and G.~Mergen, ``{M}ultipacket {R}eception in {R}andom
  {A}ccess {W}ireless {N}etworks: {F}rom {S}ignal {P}rocessing to {O}ptimal
  {M}edium {A}ccess {C}ontrol,'' \emph{IEEE Commun. Mag.}, vol.~39, no.~11, pp.
  108?--112, Nov. 2001.

\bibitem{nazer11compforw}
B.~Nazer and M.~Gastpar, ``{C}ompute-and-forward: {H}arnessing {I}nterference
  {T}hrough {S}tructured {C}odes,'' \emph{{IEEE} Trans. Inf. Theory}, vol.~57,
  no.~10, pp. 6463--6486, Oct. 2011.

\bibitem{MACbook}
D.~Danyev, B.~Laczay, and M.~Ruszinko,
  ``\BIBforeignlanguage{English}{{M}ultiple {A}ccess {A}dder {C}hannel},'' in
  \emph{\BIBforeignlanguage{English}{Multiple Access Channels}}, E.~Biglieri
  and L.~Gyorfi, Eds.\hskip 1em plus 0.5em minus 0.4em\relax IOS press, 2007,
  pp. 26--53.

\bibitem{lindstrom1975}
B.~Lindstr\"{o}m, ``Determining subsets by unramified experiments,'' in \emph{A
  Survey of Statistical Design and Linear Models}, J.~N. Srivastava, Ed.\hskip
  1em plus 0.5em minus 0.4em\relax North-Holland, 1975.

\bibitem{tsybakov1979ergodicity}
B.~S. Tsybakov and V.~A. Mikhailov, ``{E}rgodicity of a {S}lotted {ALOHA}
  {S}ystem,'' \emph{Proble. Pered. Inform.}, vol.~15, no.~4, pp. 73--87, 1979.

\bibitem{tsitsiklis86report}
J.~N. Tsitsiklis, ``Analysis of a {M}ultiaccess {C}ontrol {S}cheme,'' MIT,
  Tech. Rep. LIDS-P-1534, 1986.

\end{thebibliography}


\end{document}